# LEARNING REPOSITORY ADAPTABILITY IN AN AGENT-BASED UNIVERSITY ENVIRONMENT

Vanco Cabukovski[1], Roman Golubovski[1] and Riste Temjanovski[2]

[1]Faculty of Natural Sciences and Mathematics, Ss Cyril and Methodius University, Skopje, Macedonia

[2]Faculty of Economics, "Goce Delcev" University, Stip, Macedonia

## ABSTRACT

*Automated e-Learning Systems (AeLS) are fundamental to contemporary educational concepts worldwide. It has become a standard not only in support to the formal curriculum, but containing social platform capabilities, gamification elements and functionalities fostering communities of experts, also for faster knowledge dissemination. Additionally, AeLSs support internal communications and customizable analytics and methodologies to quickly identify learning performance, which in turn can be used as feedback to implement adaptability in tailoring the content management to meet specific individual needs. The volume of fast growing AeLS content of supplement material and exchanged communication combined with the already huge material archived in the university libraries is enormous and needs sophisticated managing through electronic repositories. Such integration of content management systems (CMS) present challenges which can be solved optimally with the use of distributed management implemented through agent-based systems. This paper depicts a successful implementation of an Integrated Intelligent Agent Based University Information System (IABUIS).*

## KEYWORDS

*Intelligent University Information System; Library Information System; Adaptive e-Learning System.*

## 1. INTRODUCTION

Agent-based systems technology has generated lots of excitement in the recent years because of its promise as a new paradigm for conceptualizing, designing and implementing software systems. The high majority of agent-based systems consist of a single agent. However, as technology gets more advanced, addressing complex applications increases and the need for distributed systems consisting of multiple agents that communicate in a peer-to-peer fashion is becoming apparent [1][2].

Multi-agent systems are designed as a collection of interacting autonomous agents, each having their own capacities and goals that are situated in a common environment. This interaction might involve communication, i.e. passing information, from one agent to another and with their environment [2].

An information system is very important part of the contemporary university. The university achievements in the education and the science areas straightly depend on the university activities computerization and the level of that computerization [3]. The university information system is created as an integrated system, aiming to computerize all university processes. Processes use general university registers, general classifications. It allows for individual users to avoid the different interpretation of the same data. All data are stored in a centralized database. This way it allows avoidance of data duplication. Information is gathered, stored in a database and managed, where the first sources are placed, i.e. in the university subdivisions. That's why the circulation time of documents decreases, less logic mistakes are made, and search and elimination of mistakes can be done effectively. The information system is based on University computer network. Fiber optic lines connect the most university subdivisions. It ensures quick and reliable information transmission. The client–server architecture is used for the university information system. It enhances data reliability and security, in connection with Internet technologies.

Professors and university staff have many different type of work to prepare. University resources (time, money, classrooms etc.) need to be allocated in a detailed manner by the staff. Scheduling and planning of exams must be done in exact time by professors. Huge production of teaching material each year, administrative and teaching data and library's collections need to be integrated on university level in knowledge and information university system, fully standardized and made compatible on state and global level.

A model of an Integrated Intelligent (Agent-Based) Univesity Information System - IABUIS with an embedded multi-agent infrastructure has been developed at the Faculty of Natural Sciences and Mathematics with the University Ss. Cyril and Methodius. It is designed to control University's administration and education system. Many procedures related with educational and non-educational programs in an university environment are supported by the outcomes of this project. With the very first steps in development of this system one could be introduced in [4]. The IABUIS main structure was described in [3][5] and [6].

The IABUIS includes four very important modules: Student Administration Management System (SAMS), Library Information System (LIS), Distance Learning System (DLS) and University Management Information System (UMIS). It is an integrated intelligent e-university environment in a provision of multi-agent infrastructure, agent-based e-learning concepts, technology and digital content unification, digital library's standardization and information management integration. Some of these aspects are discussed in [1][2][7][8].

Adaptation is the new trend in the modern e-Learning concepts (Adaptive e-Learning System - AeLS) aimed to produce more effective learning curve by tailoring a course's curriculum to individuals' specific preferences. These individual student preferences are evaluated by the AeLS in an automated manner, by following student's activities in the formal e-Learning system (lecturing material) and in the informal complementary Content Management System (CMS) containing carefully gathered supplement material (multimedia supplements).

An AeLS is able to keep track of individual usage and to accommodate content automatically for each of the users, for the best learning result, which in turn is supported by a student model built from student's goals, preferences, and knowledge. Then the student model is used to adapt the interaction mode of the e-Learning system according to the user's needs [9][10].

An AeLS can enhance the usability of the learning material by addition of more effective supplements of fellow students and lecturers' knowledge dissemination, which have proved to improve the student acquisition lead-in to better learning and examination results. The adaptive behaviour of a learning environment has numerous manifestations: adaptive interaction, adaptive content/course delivery, content discovery and assembly, and adaptive collaboration support.
The Adaptive Content/Course Delivery applied to IABUIS and the DLS constitutes the most common and widely used collection of adaptation techniques applied in learning environments today. It optimises the fit between course contents and user characteristics / requirements, so that the optimal learning result is obtained. The AeLS is based on adaptive selection of alternative supplements of course material. A methodology for development of additional low-budget digital content as an additional alternative fragments of course material for adaptive selection and composing of the course to individual users in accordance with their knowledge and behaviour, the influence of the low-budget digital contents in the increase of the quality of the students' experience, and the cost analysis of the additional digital content are given in [11].

In this paper, presented is an AeLS successfully implemented at the Faculty of Natural Sciences and Mathematics with the University Ss. Cyril and Methodius, as an advancement from the previous agent-based eLS IABUIS (Integrated Intelligent Agent Based University Information System). The main point of interest would be an ULIS adaptability as a part of the AeLS (Adaptive e-Learning System) which attempts to propagate faster individual learning curves by employing agent-based system consisted of agent-based algorithms for adaptive interaction with the consumers (students), and adaptive content/course selection and delivery of appropriate material (supplements) intended for improved knowledge acquisition, thus better learning results - subject of official examination. An AeLS ULIS adaptability is developed based on adaptive selection of alternative (fragments of) course material developed as low-budget content with a main goal alternatively to give the students additional material in order to successfully gain and implement the knowledge as well to automatically suggest literature from the ULIS.

Such an adaptive eLS is by definition demanding of intellectual and material effort, directly implicating round-the-clock maintenance. Therefore, one of the crucial factors in its conceptualizing, dimensioning and realization is the financial one. Making of supplements (often synchronized video and audio content) would normally be expensive. However, required technologies are already affordable at little or no cost at all - usually open source or freeware, thus allowing for cheap low-budget multimedia production. These circumstances support the CMS of the AeLS to be continuously updated and maintained according to the trended requirements. This kind of low-budget CMS shows almost immediately efficiency and usefulness in supporting students' advancement in their studies through successful overcoming the learning curriculum in both the theoretical as well as the practical courses.

In Section 2, the model of IABUIS with an embedded multi-agent infrastructure will be presented. In Section 3, an ULIS (University Library Information System) is described with a learning repository implementation. In Section 4 an ULIS adaptability and its cooperation with the IABUIS intelligent software agents is given. This extension in a manner of ULIS adaptibility is an essential part of a Distance Learning System (DLS) and IABUIS.

## 2. THE IABUIS MODEL

The model of IABUIS includes four processes: student administrative information management, library information management, e-learning information management and university administrative information management process. In Fig. 1, the communication of these processes with the "external world" is given.

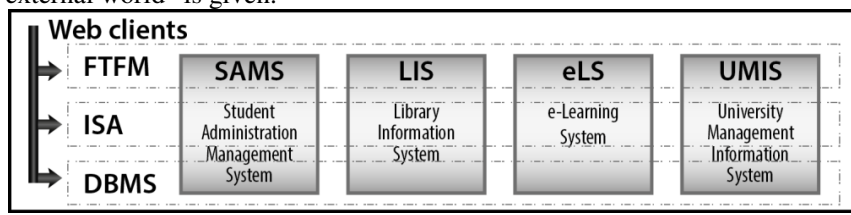

Figure 1. Communication of the IABUIS processes with the "external world"

Special web-based segments serve the connection with the IABUIS elements: the free text, files and multimedia management system (FTFM); the data base management system (DBMS) and the independent software applications (ISA).

In our model, each subdivision (institute or department) is represented by an Intranet, while all subdivisions together constitute the University Intranet. The University Intranet with a few external connections consist the University Extranet [5][12], which is connected to the Internet on several points.

The FTFM management system is consisted of portals, www pages, questionnaires, forms, output reports, images, servers (like e-mail and ftp), etc. The DBMS consists of databases, and advanced indexing and retrieval database engines (ORACLE and MySQL). The third segment (ISA) contains different software applications developed for special purposes, like accounting, warehousing, book circulation, optimal schedule of classes, distance learning management systems, etc.

The Student Administration Management System (SAMS) keeps track of the students educational records. It consists of the following modules: Academic advisement for tracking the requirements and regulations that a student must satisfy in order to graduate, Admissions, Student Financials, Student records and Campus community. This system can be accessed by students to process enrollment transactions, change addresses, and view academic records. The university administrators can access the system to maintain Student Educational Records.

The Library Information System (LIS) is developed in ORACLE. It is UNICODE based and it is supporting the MARC format and the dialects UNIMARC and MARC21. The system is composed of the following modules: Requisition, Cataloguing, Transactions, Library fund, Searching, Updating, System administration, Report and Help. The process of cataloguing is centralized. The access for the public is via Internet. This module would be described in a next section.

The Distance Learning System (DLS) is based on a ORACLE iLearning management system providing complete infrastructure to manage, deliver, and track training in both online and classroom based environments. It supports different learning styles, from self-paced material, to scheduled or synchronous classes offline or online and collaborative and integrated learning.
The University Management Information System (UMIS) is designed to enhance the efficiency of the administrative and managerial aspects of the institution. It consists of the following modules: Human resources, Accounting, Finances, Planning and Document management and archiving. Stored data is organized in unique ways that empower the administrators and university management to make better decisions. The system is fully web based supporting the principle of 'anytime anywhere'.

## 3. THE UNIVERSITY LIBRARY INFORMATION SYSTEM (ULIS)

Globalization is a very real phenomenon that is transforming information services and the library systems into e-library, digital library and semantic digital library according to the evolution of information technology. The library information systems in present day have rapidly evolved into the digital library aiming to realize integration and interoperability of information resources under distributed computing environment based on the Internet and computer networks. This kind of the transition of library information systems is to accommodate new library resources due to the innovation of information technology, neither the users' requirements nor the new service support. A new market for structured information, knowledge and skills needed metadata services. Metadata or structured data needs to be easily transferrable between different users and institutions, and satisfy consumer "starving" for information. Today's library users expect speed and immediacy of information discovery, one-stop access to aggregated services, user-generated open content, and personalised, workflow-related delivery to the desktop.

Since the main business objects of the library are knowledge resources, the library information systems are heavily influenced by means of the evolution of information. On the proliferation of electronic resources such as audio, images, videos and texts, electronic library (e-Library) appear to manage electronic resources effectively. [13]

The ULIS which is part of IABUIS is standardized librarian information system, intended to catalogue, update, search, borrow book and non-book materials (artwork, audio recordings, video recordings, cartography) in academic, school, popular and public libraries, archives, museums, film archives, institutions with other types of needs. This complex information system, also can perform automation of daily operations as well as documents archives of ministries, government offices and public enterprises who need a modern and powerful UNICODE based (multilingual) library system compatible with the standard MARC and its dialects UNIMARC, MARC21 and others. ULIS is based on ORACLE database engine, fully secure and protected system with the possibility of networking and web access to the data.

The content organized by the ULIS is "merged" with the growing CMS hosting multimedia supplements to the existing courses into the integral learning Repository. The Repository is an effective system for maintaining the huge volume of fast growing collection of educational items belonging to the ULIS (bookware and courseware) and the multimedia CMS (low-budget video tutorials). The Repository is conveniently interfaced with the rest of the IABUIS and managed by the distributed multi-agent system which suggests students material from both contents adaptively.

Despite the full compatibility with other library systems based on the MARC standard - an internationally accepted standard for bibliographic / catalog processing of library material, ULIS offers and the following:

- Multilanguage based on UNICODE standard text-editor for the full support of all the world's languages and their specifics.
- Automation of procurement of new library materials (requests for procurement, seeking approval, approval of procurement, order procurement acceptance, financial records).
- Automated Cataloging / monographs, serials, non-book materials, articles, books / signing, enumeration and recording, inventory and book of records, indexing, annotation, the main catalog card, analytical catalog card UDC catalogs ABCDE catalogs, bibliographies .
- Automation of the circulation of library materials (lending, return, reservation loss, withdrawal of damaged samples, interlibrary loans, financial records).
- Automatic control and review the state of the library; Generating periodic reports and statistical surveys.
- Basic and advanced searching local databases supported by the Z39.50 / ISO 23950 international standard;
- Classification by UDC, DEWEY other world classification schemes.
- Develop / update auxiliary files.
- WEB access to library materials (search, book, review of reserved and borrowed books).
- Import / Export of UNIMARC records in standard ISO 2709.
- Import / Export of records in MARC other internationally accepted standards.

The standards implemented in ULIS are:

- MARC standard, internationally accepted standard for bibliographic / catalog Processing librarian material and its dialects UNIMARC, MARC21 and others.
- UNICODE standard for multilingualism;
- ISBN, ISSN, UDC, DEWEY, Z39.50, ISO 2709 and other internationally accepted standards for library processes functioning;
- Inventory standards;
- implemented quickly and efficiently (simple and advanced) search engine as the very important aspect of the librarian operation;
- Multilingualism included in the BIS provides the ability to search in all languages.

The web and its associated technical standards continue to dominate, although within a framework of much more use of mobile devices, data protection take primarily place in system's platform. ULIS covers such limited access to the data protecting their integrity in all domains and sources to access. In the next time, ULIS will be moving towards cloud computing technology and taking advantages of cloud based services especially in building digital libraries, social networking and information communication with manifold flexibilities but some issues related to security, privacy, trustworthiness and legal issues were still not fully resolved.

The use of cloud based library management systems has increased drastically since the rise of "cloud" technology started. Cloud computing is acting as a resources pooling technology for accessing infinite computing services and resources as per demand of users and can be compare with models of pay as you use or utility model same as used for mobile services usages and electricity consumption. Cloud based services provide a means for libraries to free resources on information technologies and focus on libraries' core competencies- manage, organize and disseminate information. Cloud based services are also bringing cutting-edge services to libraries that have less information technology expertise [14][15].

## 4. THE ULIS ADAPTIBILITY AND THE IABUIS AGENTS COOPERATION

The eLS had successfully been upgraded from a classic e-Learning System to an Adaptive eLS (AeLS) by employing the multi-agent environment of the IABUIS aiming to improve the educational process.

The system presented is based on adaptive selection of additional learning material (books, articles and video tutorials) stored in the Repository to the individual student needs This AeLS is successfully implemented as an advancement from the previous agent-based eLS IABUIS. The general idea behind this concept is to provide the students with supplemental material in support to steeper learning curves, submitted to the AeLS via the Repository CMS (RCMS), and approved by the lecturers. The RCMS is constantly monitoring through SAMS the overall progress of all students with their semestral courses and updates the supplements ranking list accordingly. Students are also evaluated periodically through test examination and ranked accordingly. Adaptation is then implemented with an algorithm which basically suggests higher ranked supplements set to lower ranking students, and more relaxed content to higher ranking students.

In addition to the 'informal' multimedia supplements, the AeLS implemented within the IABUIS had recently been integrated through its RCMS with the ULIS - the library being the source of adequate and recommended text books and related scientific journals/articles. This new quality required structural and functional rearrangement of the multi-agent driving the follow-up of both - the individual students progress (based on SAMS data) as well as the appropriateness of the suggested learning material by the AeLS (within the eLS). The logical integration of the CMS and the ULIS is organized through a complex DBMS based Repository (the RCMS) consisted of

International Journal of Computer Science & Information Technology (IJCSIT) Vol 8, No 3, June 2016

the relevant portions of the both, and suitable software is developed to interface them to the users and agents/applications.

By integrating RCMS with the IABUIS the AeLS is able to implement its proven agent strategies facilitating the required adaptation of the learning material, consisted of both the more formal text books and the less formal low-budget supplements.

The Adaptive Content / Course Delivery applied to the system presented in this paper constitutes the most common and widely used collection of adaptation techniques applied in learning environments today. It optimises the fit between course contents and user characteristics / requirements, so that the optimal learning result is obtained. As per Brusilovsky (2001) [18] the most typical examples of adaptations in this category are: dynamic course (re-) structuring; adaptive navigation support; and, adaptive selection of alternative supplements. Our system presented is based on adaptive selection of alternative supplements of course material.
A methodology for development of additional low-budget digital content as an additional alternative fragments of course material for adaptive selection and composing of the course to individual users in accordance with their knowledge and behaviour, the influence of the low-budget digital contents in the increase of the quality of the students' experience, and the cost analysis of the additional digital content are given in [11].

In this paper, presented is an AeLS successfully implemented at the Faculty of Natural Sciences and Mathematics with the University Ss. Cyril and Methodius, as an advancement from the previous agent-based eLS IABUIS (Integrated Intelligent Agent Based University Information System). The main point of interest would be an AeLS (Adaptive e-Learning System) within it which attempts to propagate faster individual learning curves by employing agent-based system consisted of agent-based algorithms for adaptive interaction with the consumers (students), and adaptive content/course selection and delivery of appropriate material (supplements, text books and journal papers/articles) intended for improved knowledge acquisition, thus better learning results - subject of official examination. An AeLS is developed based on adaptive selection of alternative course material developed as low-budget or library content with a main goal alternatively to give the students additional material in order successfully to gain and implement the knowledge. In the paper presented are the latest improvements in the Personal Filtering Assistant agent, the user - AeLS interaction point.

The structure of the updated AeLS with learning repository (RCMS) and the embedded agent infrastructure is presented with the following model given in figure 2.

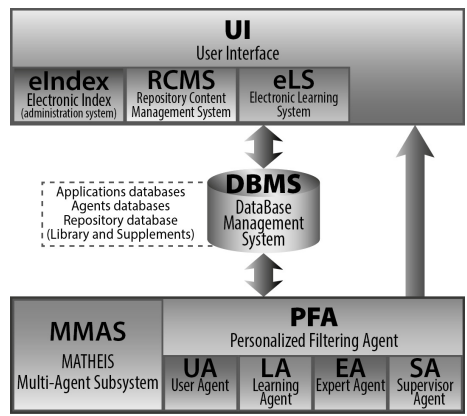

Figure 2. The AeLS structure

International Journal of Computer Science & Information Technology (IJCSIT) Vol 8, No 3, June 2016

Depicted AeLS is an adaptive agent based extension to the existing distance educational system MATHEIS (MATHematical Electronic Interactive System). MATHEIS is an educational system for learning mathematics and informatics for pupils and students [4][16]. This system has been successfully integrated into ORACLE iLearning management system at the Faculty of Natural Sciences and Mathematics in Skopje and extended into Sakai e-learning environment. And a CMS (Content Management System) is integrated with the two, to host multimedia supplements to existing courses, along with the University Library - ULIS. The learning Repository (RCMS) is the newest DBMS organization of the administrative applications' data, the low-budget supplements' CMS and the ULIS archives.

An agent extension of MATHEIS empowers the system with the following features: monitoring the students behaviour and interests at the system; determining the student's skill level; enabling cooperative task resolution among students; enabling different views of the services and content according to the student's skills and requirements (adaptability); notifying the students when the newest tests for appropriate level are available; presenting the tests to the students and estimating the received results; automatically updating of the student's levels depending on the estimated results, etc..

The general idea behind this concept is to provide the students with supplemental material in support to steeper learning curves, submitted to the AeLS via the CMS, and approved by the lecturers. The CMS is constantly monitoring the overall progress of all students with their semestral courses and updates the supplements ranking list accordingly. Students are also evaluated periodically through test examination and ranked accordingly. Adaptation is then implemented with an algorithm which basically suggests higher ranked supplements set to lower ranking students, and more relaxed content to higher ranking students. This adaptive aspect of the eLS (AeLS) is implemented by the PFA subsystem.

**4.1 THE MMAS SUBSYSTEM**

The agent-based structure of the AeLS core (MATHEIS Multi-Agent Subsystem) is given in figure 3.

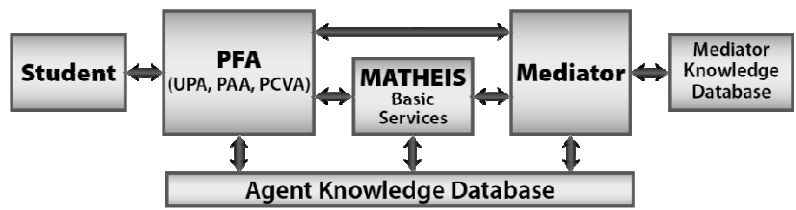

Figure 3. The agent-based structure of MATHEIS

The system is able to assist adaptively in filtering of the educational material according to the UPA (User Profile Agent) and student's activities in the communication with MATHEIS basic services recorded by the PAA (Personalized Activity Agent). The Personalized Content Viewing Agent (PCVA) is responsible for the adaptive interaction and adaptive content/course delivery – this enhances the usability of material and thus make the e-Learning system more effective, which improves the students' acquisition of knowledge and lead to better learning results. All three agents collaborate among them sharing distributed agent knowledge and learning rules [1][2].

The Mediator is responsible for the student learning model, database of the student's grades, degree levels, preferences, abilities, aptitudes, etc. This agent communicates with the Mediator knowledge database, generated during the educational process on the system.

The Personalized Filtering Assistant (PFA) segment follows the student's activities. It is responsible for the adaptive selection and display of the content (of both supplements and text books/articles) and adaptive interaction. The PFA is trained for each student to make the right content selection appropriate to the student's abilities and aptitudes. Since the student's abilities and aptitudes are not assumed to be constant overtime, the system is able via pre-selected tests to notice that the student's abilities and aptitudes have been changed. The system adapts its behaviour in response to these changes. Any new student on the system, after few pre-selected tests for evaluation of grade level and estimation of student's abilities and aptitudes will be accompanied with appropriate personal set of agents adjusted to the student's knowledge and interests.

The functional structure of MMAS as shown in figure 4 is consisted of four conceptual subsystems: The User Agents Community; the Level Maintenance Subsystem; the Supervisory Subsystem and the Fuzzy Expert Subsystem. For detailed description of the structure of MMAS as well as relations between agents in MMAS (read: Cabukovski 2010a). This paper emphasizes the Fuzzy Expertize of the PFA.

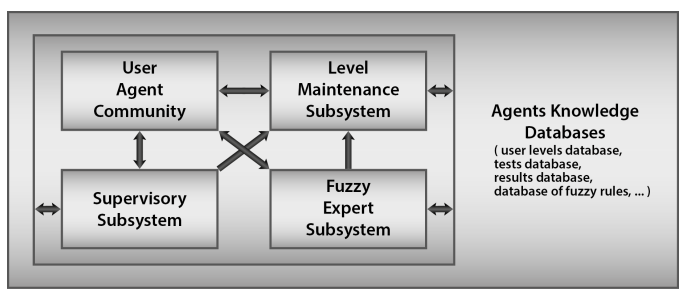

Figure 4. Relationship among the MMAS functional components

The multi-agent environment (MMAS - MATHEIS Multi-Agent System) is based on a former eLS called MATHEIS (MATHematical Electronic Interactive System) - an educational system for learning mathematics and informatics for pupils and students [4][16][17]. This system had been successfully integrated into ORACLE iLearning management system at the Faculty of Natural Sciences and Mathematics in Skopje and extended into SAKAI e-learning environment.
This adaptive aspect of the eLS (AeLS) is implemented by the PFA agent subsystem which follows the student's activities. It is responsible for the adaptive selection and display of the content (of both supplements and text books/articles) and adaptive interaction. The PFA is trained for each student to make the right content selection appropriate to the student's abilities and aptitudes.

The fundamental task of the PFA agent is to continuously evaluate both, the individual knowledge level of each student as (s)he advances through the course's curricula on one hand, as well as the "most appropriate" supplement material that had proved to be most helpful for that particular student level, on the other hand. So basically, parallel ranking lists of students (by level) and ULIS supplements (by significance) are maintained and used for matching against each other, in direct support of the adaptation process itself.

The ULIS just like any other DBMS software archives every single action within its working domain in detailed log files containing: the action itself (upload, download, approved, etc.); the user who performed it (student, lecturer); the content (supplement, book, article) affected; timestamp of the action, etc. These detailed logs allow for the PFA to keep track of what content had been downloaded by which student within a certain course. PFA can also use the SAMS system to keep track of which students had passed certain exams and with what grades. By combining these DB entries with the ULIS logs the PFA can easily calculate a precise contribution of every single supplement to students' success of its corresponding course. Calculated contribution of all pieces of content belonging to same course allows the PFA to build and maintain a content's list ranking highest those that contribute most, i.e. downloads that helped most students to pass examination and/or with highest grades. The rank lists for all courses are updated on semestral basis - after the semestral exams. Adaptation is then implemented with an algorithm which basically suggests higher ranked supplements set to lower ranking students, and more relaxed content to higher ranking students.

The PFA agent follows all student's activities and among other things performs the adaptive selection of supplements to suggest to individual students for particular course. In order to be able to adapt supplements' curricula to the student, the PFA has to evaluate student's skill level periodically by test examination. The nature of the level estimation process logically requires a fuzzy approach, which is the job of the Fuzzy Expert System.

### 4.2 THE PERSONAL FILTERING ASSISTANT AGENT

The adaptive behaviour of a learning environment is implemented in the PFA agent, and has numerous manifestations. The broad and partially overlapping categories are: adaptive interaction, adaptive content/course delivery, content discovery and assembly, and adaptive collaboration support.

The Adaptive Interaction applied to our system, refers to adaptations that take place at the system's interface and are intended to facilitate or support the user's interaction with the system, without, however, modifying in any way the learning content itself. The adaptations at this level include: the employment of alternative graphical or colour schemes, font sizes, etc., to accommodate user preferences, requirements or abilities at the lexical (or physical) level of interaction; the reorganization or restructuring of interactive tasks at the syntactic level of interaction; or the adoption of alternative interaction metaphors at the semantic level of interaction.

The fundamental task of the PFA agent is to continuously evaluate both, the individual knowledge level of each student as (s)he advances through the course's curricula on one hand, as well as the "most appropriate" supplement material that had proved to be most helpful for that particular student level, on the other hand.

So basically, parallel ranking lists of students (by level) and supplements (by significance) are maintained and used for matching against each other, in direct support of the adaptation process itself.

### 4.3 THE ROLE OF THE CMS AND THE ULIS LEARNING REPOSITORY

The supplements (portions of particular course supporting/related content) are multimedia units managed in the Repository Database (DBMS) by a Content Management System (CMS). The publications (text books, journals/proceedings articles and papers) residing in the same Repository DBMS are managed by the ULIS. The RCMS is integrated in the faculty's AeLS with its own UI through which its portion of the integrated DB is updated and maintained. Anybody can submit a supplement, however only a course corresponding lecturer can admit it for usage after reviewing it. The RCMS also manages the ULIS archives.

The RCMS just like any other DBMS software archives every single action within its working domain in detailed log files containing: the action itself (upload, download, approved, etc.); the user who performed it (student, lecturer); the content (supplement, book, article) affected;

timestamp of the action, etc. These detailed logs allow for the PFA to keep track of what content had been downloaded by which student within a certain course. PFA can also use the eIndex system to keep track of which students had passed certain exams and with what grades. By combining these DB entries with the CMS logs the PFA can easily calculate a precise contribution of every single supplement to students' success of its corresponding course. The RCMS can also retrieve same logs from the ULIS.

Several parameters can be used to evaluate a content's rank on a cumulative basis:
- Total number of students that have downloaded the content
- Total number of the above that have passed the exam
- The number of students that have downloaded it after being suggested to
- The number of students that have downloaded it without being suggested
- Average students grade
- Average students that downloaded the supplement grade
- etc.

Calculated contribution of all pieces of content belonging to same course allows the PFA to build and maintain a content's list ranking highest those that contribute most, i.e. downloads that helped most students to pass examination and/or with highest grades. The rank lists for all courses are updated on semestral basis - after the semestral exams.

As being said above and explained in the next section, students are also evaluated periodically through test examination and ranked accordingly. Adaptation is then implemented with an algorithm which basically suggests higher ranked supplements set to lower ranking students, and more relaxed content to higher ranking students.

### 4.4 THE ROLE OF THE FUZZY EXPERT SYSTEM

The PFA agent follows all student's activities and among other things performs the adaptive selection of supplements to suggest to individual students for particular course. In order to be able to adapt supplements' curricula to the student, the PFA has to evaluate student's skill level periodically by test examination. The nature of the level estimation process logically requires a fuzzy approach, which is the job of the Fuzzy Expert System (FES).

Namely, in the problem solving process not only the final result is important, but also all intermediate stages should be treated as well. Thus, often is not enough to esteem the test solution as true or false, but also partial truth of the solution needs be considered. Additional problem is the different ways of representing same result Things become even more complicated when tests are consisted of more than single task, so the estimation of the test depend on all tasks' solutions. These issues are important because determination of user's skill level relays on all of them.

Fuzzy logic is particularly applicable to the individual student level estimation since problems of partial truth are very common in the real world. It is a kind of generalization of the two-element Boolean logic with values between 0 and 1. Fuzzy subset P of a set S is defined as mapping $f : S \rightarrow [0,1]$ where the second element of the ordered pairs $(x,t)$: $x \in S$, $t \in [0,1]$ obtained from the mapping represents the degree of membership in the subset P, or the degree of truth of the statement "x is in P". The set S is referred to as universe of discourse, and f is a membership function of P.

In the PFA context, the universe of discourse S is consisted of tests T for user skills estimation. A test is represented as a questionnaire of $n$ tasks, where for each task several solutions are proposed, only one of which is correct. This means that a test can be viewed as an $n$-dimensional vector of tasks T=$(t_1, ..., t_n)$, and the solution of the test as an n-dimensional vector of results R=$(r_1, ..., r_n)$, where $r_i \in \{0,1\}$ and $r_i$=1 means correct user reply and $r_i$=0 means incorrect result. For every task in the test a weighting factor $k$ is assigned, where the condition $k_1+k_2+ ... +k_n = n$ must be satisfied. Without losing generality it can be assumed that $m$ correct results ($m \leq n$) appear at the first m places in both vectors T and R. Two membership functions can be defined:

$$solved(T) = (k_1 + k_2 + ... + k_m) / n , mistaken(T) = 1 - solved(T)$$

where T∈S, $m$ is the number of accurate results, $k_i$ ($1 \leq i \leq m$) is the weighting factor of the $i$-th task, and $n$ is the number of tasks.

For the level computation two membership functions are also defined:

$$high(L) = L, low(L) = 1–L,$$

where L∈[0,1]. Only two simple rules will be presented in the knowledgebase of the FES:
rule R1 : if t is *solved* then l is *high* and rule R2 : if t is *mistaken* then l is *low*
where t is an input variable taking values from S, and l is the output variable.

After the fuzzification process, i.e. applying the *solved* and *mistaken* functions on a test t done by the user, the degree of truth for the input variable t is computed. Afterwards, PRODUCT-SUM inference is applied on both rules so that two output fuzzy subsets for the output variable l are obtained. The PRODUCT-SUM inference is a method of applying the truth value of the premises of every rule to the conclusion part of the corresponding rule, using the PRODUCT function for scaling the output membership function of every rule with the computed truth value of the corresponding premise, and finally applying the SUM function to all output fuzzy subsets assigned to the output variable l, in order to obtain a single output fuzzy subset for l. At the end, a defuzzification is applied on the fuzzy output subset assigned to l, in order to obtain a single crisp value representing the fuzzy subset. The Maximum method is used for defuzzification. The fuzzy logic controller first identifies the scaled membership function with the greatest degree of membership and then determines the typical numerical value for that membership function.

## 5. CONCLUSIONS

An Integrated Intelligent (Agent-Based) University Information System (IABUIS) consisted of the default administrative modules (among which the student SAMS) as well as of the formal e-Learning System (eLS) is upgraded with an adaptation functionality in support to steeper student learning curves. The novel integrated system can be considered as an Adaptive e-Learning System (AeLS). And it utilizes the rich and comprehensive University Library Information System (ULIS).

The structure of the AeLS is now consisted of the following four - an eIndex (for student files administration), an eLS (for formal course coverage with learning material), an additionally integrated CMS (Content Management System) with a Repository hosting additional low-budget multimedia supplements and its connection to the existing ULIS (University Library Information System) for navigation within the achieved text books, journals, proceedings, and related material. On data-level they all share an integrated Database.

The AeLS introduces an adaptive approach to every individual student, by inferring which supplement (unit or set) is most suitable to support his/her further learning activities. Namely, the core of the AeLS is an existing agent-based interactive context (MMAS) which defines additional roles to the Expert Agent to perform: ranking of the available supplements and library content

<em>International Journal of Computer Science & Information Technology (IJCSIT) Vol 8, No 3, June 2016</em>

within a certain course by determining their individual impact (and thus usefulness); fuzzy-based ranking of the students currently enrolled in the course; and configurable (possibly fuzzy-based) matching of the two in order to perform the best possible suggestion of supplement content to every single student to ensure steeper learning curve.

All the fuzzy processing and inference is performed by the PFA module and already shows positive results and acceptance by the student community. Further advancement is being worked on and expected as improvement in the adaptation algorithm, as well as in the fuzzy reasoning for the suggestions inference.

ULIS is used as knowledgebase supporting the AeLS in providing tailored content (books, journals, proceedings, and related material) to the students. Most of the research was focused on the seamless integration of ULIS with the supplements CMS into the complex Repository CMS (RCMS), as well as on the development of the Fuzzy based methodology for implementation of the adaptability of the RCMS. Special effort was put into recognition of specific needs for individual courses resulting in production of low-budget multimedia tutorial supplements.
However, fundamental attention was given to the MMAS infrastructure resulting from carefully interfaced agents' algorithms, allowing sophisticated management of the corresponding databases and the massive Repository of valuable learning educational material.

Along with the ongoing activities focused on technical improvement of different aspects of the AeLS, significant effort is made toward acquiring measurable feedback of its usefulness. Preliminary analysis of this system's feedback is made by tracking particular courses that are subject to continuous technological update (programming, etc.). Results are still acquired and gathered, however preliminary statistics already show visible advancement in students' grasp of the learning matter and therefore better examination results. In particular, this is noticeable in students accepting popular learning books suggestions from the ULIS, as well as those helping themselves with the supplementary multimedia tutorials. We expect an accurate and comprehensive picture on this after several semesters.

International Journal of Computer Science & Information Technology (IJCSIT) Vol 8, No 3, June 2016

**Authors**

**Vanco E. Cabukovski** is Full Professor of Software Engineering at Faculty of Natural Sciences and Mathematics, Sts. Cyril and Methodius University in Skopje, Republic of Macedonia. His main research interests include intelligent systems, e-learning systems and information systems. He has published over 70 scientific papers and over 25 books in informatics and ICT. He is author of 30 software applications and has participated in over 20 domestic and international projects.

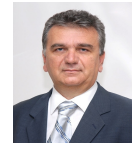

**Roman V. Golubovski** is an Assistant Professor of Software Engineering with the Faculty of Natural Sciences and Mathematics, Ss. Cyril and Methodius University in Skopje, Republic of Macedonia. His main research interests are focused on signal acquisition and processing, computerized automation and information systems, and intelligent systems. Assist. Professor Golubovski is author and co-author of 2 books and more than 33 papers. He is an author of 7 software and hardware development project applications in various areas, and participant in more than 50 application projects as well as in 4 domestic and international research projects.

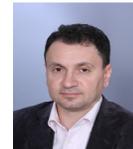

**Riste C. Temjanovski** is Full Professor of Economy of Transport Systems at Faculty of Economics, "Goce Delcev" University - Stip, Republic of Macedonia. His current research interests include transport systems and models, GIS systems, e-business and consumer behaviour analysis, experimental economics and strategy in ICT markets and entrepreneurship. Research work has been published in over 60 research articles in academic journals, 10 academic books and has participated in 4 domestic and international research projects.

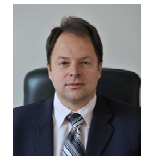